\documentclass[a4paper,11pt]{article}
\pdfoutput=1 

\usepackage{jcappub} 

\usepackage[T1]{fontenc} 
\usepackage{physics}
\usepackage{wrapfig}
\usepackage{xcolor}
\usepackage{float}
\restylefloat{table}
\usepackage{changepage}
\usepackage{booktabs}

\usepackage{csquotes}

\title{\boldmath
Influence of local structure on relic neutrino abundances and anisotropies
}

\author[a,1]{Fabian Zimmer,\note{Corresponding author.}}
\author[a]{Camila A. Correa,}
\author[a,b]{and Shin'ichiro Ando}

\affiliation[a]{GRAPPA Institute, University of Amsterdam, Science Park 904, 1098 XH Amsterdam, The Netherlands}
\affiliation[b]{Kavli Institute for the Physics and Mathematics of the Universe (WPI), University of Tokyo, Chiba 277-8583, Japan}
\emailAdd{f.zimmer@uva.nl}
\emailAdd{c.a.correa@uva.nl}
\emailAdd{s.ando@uva.nl}

\abstract{Gravitational potentials of the Milky Way and extragalactic structures can influence the propagation of the cosmic neutrino background (CNB). Of particular interest to future CNB observatories, such as PTOLEMY, is the CNB number density on Earth. In this study, we have developed a simulation framework that maps the trajectories of relic neutrinos as they move through the local gravitational environment. The potentials are based on the dark matter halos found in state-of-the-art cosmological N-body simulations, resulting in a more nuanced and realistic input than the previously employed analytical models. We find that the complex dark matter distributions, along with their dynamic evolution, influence the abundance and anisotropies of the CNB in ways unaccounted for by earlier analytical methods. Importantly, these cosmological simulations contain multiple instances of Milky Way-like halos that we employ to model a variety of gravitational landscapes. Consequently, we notice a variation in the CNB number densities that can be primarily attributed to the differences in the masses of these individual halos. For neutrino masses between $0.01$ and $0.3$~eV, we note clustering factors within the range of $1+\mathcal{O}(10^{-3})$ to $1+\mathcal{O}(1)$. Furthermore, the asymmetric nature of the underlying dark matter distributions within the halos results in not only overdense, but intriguingly, underdense regions within the full-sky anisotropy maps. Gravitational clustering appears to have a significant impact on the angular power spectra of these maps, leading to orders of magnitude more power on smaller scales beyond multipoles of $\ell = 3$ when juxtaposed against predictions by primordial fluctuations. We discuss how our results reshape our understanding of relic neutrino clustering and how this might affect observability of future CNB observatories such as PTOLEMY.

\vspace*{5pt} \noindent \textbf{\texttt{GitHub}}: Our simulation code will be made visible \href{https://github.com/Fabian-Zimmer/neutrino_clustering.git}{here}.
}


\begin{document}
\maketitle
\flushbottom

\section{Introduction}
\label{sec:intro}
The cosmic neutrino background (CNB) is one of the last fundamental predictions of the cosmological model which remains undetected in a laboratory setting. Indirect evidence was found as phase shifts in the cosmic microwave background (CMB) and baryon acoustic oscillations (BAO) power spectra~\cite{Follin:2015hya,Baumann:2019keh}. Additionally, the precise determinations of the effective number of relativistic species in the early Universe from the Planck collaboration~\cite{Planck:2018vyg}, from big bang nucleosynthesis (BBN) analyses~\cite{Pisanti:2020efz,Fields:2019pfx} and the theoretical predictions of its value from the $\Lambda$CDM cosmological model~\cite{Akita:2020szl,EscuderoAbenza:2020cmq,Bennett:2019ewm,Froustey:2020mcq,Cielo:2023bqp} all are in remarkable agreement and give us high confidence of the existence of three neutrino families in the early Universe. Understanding the subsequent evolution of the neutrinos comprising the CNB, henceforth referred to as relic neutrinos, is subject to ongoing theoretical and experimental efforts. Of vital importance to future experiments aiming to detect relic neutrinos, especially for neutrino capture on beta-decaying nuclei experiments such as PTOLEMY \cite{Betts:2013uya,Long:2014zva,PTOLEMY:2019hkd}, is the number of relic neutrinos in the vicinity of Earth. We refer to this as the local number density $n_\nu$. The total (unpolarized) capture rate for this experiment is $\Gamma = N_T n_\nu \langle \sigma v_\nu \rangle$, where $N_T$ is the number of target nuclei (tritium in the case of PTOLEMY) in the detector and $\langle \sigma v_\nu \rangle$ is the velocity-averaged cross section for the relic neutrinos. Thus, a better understanding of the number density and velocity distribution of relic neutrinos will be needed to interpret future measurements.

Cosmological N-body simulations provide a way to study relic neutrinos in a controlled yet realistic environment. This field of cosmology is rapidly advancing thanks to the improvement in the modeling of large-scale structure and the proliferation of computational facilities, as well as an ever-increasing 
number of codes for cosmological simulations including neutrinos (see~\cite{Euclid:2022qde} for an extensive code comparison). There has been growing interest in using cosmological simulations to study a variety of properties concerning relic neutrinos, such as, e.g., their helicity composition \cite{Hernandez-Molinero:2022zoo,Hernandez-Molinero:2023jes} or the radial density profile of neutrino halos~\cite{2013JCAP...03..019V,Brandbyge:2010ge,Elbers:2020lbn}. To specifically determine the local number density one needs to accurately model the gravitational potential of the local neighbourhood (and potentially beyond), which can cause relic neutrinos to cluster and form regions of varying density. It is not yet well understood, whether the region of space containing the Milky Way, specifically Earth, has an enhanced or suppressed abundance of relic neutrinos.\footnote{An upper limit on the neutrino number density is given by KATRIN~\cite{KATRIN:2022kkv}, which is many orders of magnitudes above the cosmological average of ${\sim}56 \mathrm{cm^{-3}}$. The conceivability of such values is discussed in~\cite{Bondarenko:2023ukx}.}

Aside from using methods based on linear perturbation theory to calculate these number densities~\cite{Singh:2002de,Alvey:2021xmq,Holm:2023rml} there has been a series of studies improving the methodology and scalability as well as reducing the computational costs of relic neutrino clustering calculations~\cite{Ringwald:2004np,deSalas:2017wtt,Zhang:2017ljh,Mertsch:2019qjv}.
In these studies they reconstructed the trajectories by solving the equations of motion in terms of velocity for a specific neutrino mass. A subsequent reweighting of the obtained trajectories then enables the generalization of the results of one simulation for a specific neutrino mass to a variety of neutrino masses. The gravitational potentials and their evolution in those studies were described with analytical functions that obey either spherical or cylindrical symmetries. A key contribution of our work lies in the use of asymmetric dark matter (DM) distributions and their more dynamic time evolution to calculate the potentials. We leverage the so-called ``merger trees'' of simulated DM halos, which incorporate their accretion, merger, and mass-loss events, thereby representing more realistic timelines of MW-type DM halos.\footnote{The history of our Milky Way galaxy is believed to be filled with such accretion and merger events~\cite{2020MNRAS.498.2472K}.}

Our neutrino clustering simulation framework interweaves the specially designed methodology of~\cite{Mertsch:2019qjv} with the results of state-of-the-art large-scale N-body simulations. We use cold dark matter (CDM) simulations presented in~\cite{Correa:2022dey} as an external input for the gravitational landscape. As the main objective of this study to evaluate the impact of the DM distribution (and its evolution) of single MW-type halos on the CNB, we have extracted 27 such halos contained in a 25 Mpc cosmological box. By inheriting the non-linear dynamics and evolution of these DM halos, our simulations predict novel emergent phenomena that cannot be replicated when using previous analytical treatments for the gravitational potentials.

Besides influencing the overall number density, the DM inhomogeneities the relic neutrinos encounter en route to Earth induce anisotropies in the CNB. Polarizing tritium targets in PTOLEMY, as explored in~\cite{Lisanti:2014pqa}, would make it sensitive to the incoming direction of the neutrinos,\footnote{Another recent set of studies proposes to take advantage of the large de-Broglie wavelength of these low energy neutrinos for detection~\cite{Arvanitaki:2022oby,Arvanitaki:2023fij}, which ought to be sensitive to relic neutrino directions as well.} possibly allowing the study of CNB anisotropies in the future. Within this context, we also utilize our simulations to explore how the number densities vary across the sky and how these anisotropies compare to those resulting from evolving primordial density fluctuations computed with linear theory~\cite{Michney:2006mk,Hannestad:2009xu,Tully:2021key}. As emphasized in~\cite{Hannestad:2009xu}, linear theory fails for neutrino masses $\gtrsim 0.1$ eV, necessitating the use of N-body simulations to obtain valid CNB spectra in this regime.

The remainder of this work is structured as follows: We begin by describing the theory in section~\ref{sec:theory}, specifically showing how to compute the number densities step-by-step. Section~\ref{sec:methods} describes the methodology and framework of our simulations, and section~\ref{sec:results} presents our findings. The potential consequences of our results for CNB detection prospects are discussed in section~\ref{sec:discussion}. Finally, we conclude in section~\ref{sec:conclusions_and_outlook} and contemplate further potential applications of this simulation approach.

\section{Theory}
\label{sec:theory}

In this section we describe the order of execution for the underlying equations necessary to compute the local number density of relic neutrinos, which we adapt from \cite{Mertsch:2019qjv, Parimbelli:2020jjj}.

After neutrinos decoupled from the ${\sim}1$ MeV plasma in the early Universe, they free-streamed with relativistic velocities. Due to their massive nature, however, they transitioned to a non-relativistic regime around $z \simeq 1100 \left( m_\nu / 0.58 ~\mathrm{eV} \right)$ (see, e.g., ref.~\cite{Escudero:2020ped}).\footnote{The time of photon decoupling happened around $z \sim 1100$.} Throughout this work, we consider their initial momentum to follow a Fermi-Dirac distribution with zero chemical potential, $f(p) = [1 + \exp(p/T_{\nu})]^{-1}$. The CNB temperature today, $T_{\nu,0}$, is obtained via the usual relation to the observed CMB temperature today of $T_{\nu,0} = (4/11)^{1/3} T_{\gamma,0} \approx 1.95 \, \mathrm{K}$~\cite{Planck:2018vyg}, and it was $T_\nu = (1+z)T_{\nu,0}$ at earlier redshifts. Spanning the momentum 3 decades around the CNB temperature, i.e. $0.01 \leq p/T_{\nu,0} \leq 10$, already accounts for more than $99\%$ of the distribution. This implies, that even with the minimal mass scenario\footnote{We know the mass squared differences between the mass eigenstates to be $\Delta m^2_\mathrm{sol} \approx 7.4 \times 10^{-5} \, \mathrm{eV}^2$ and $|\Delta m^2_\mathrm{atm}| \approx 2.5 \times 10^{-3} \, \mathrm{eV}^2$ from solar and atmospheric neutrino oscillation experiments, respectively~\cite{Esteban:2020cvm}, which guarantees at least one and two massive neutrinos around $0.05$ eV for normal and inverted mass ordering, respectively, even when one state is massless.} of $\sum m_\nu \approx 58 \, (100) \, \mathrm{meV}$ for normal (inverted) mass ordering, the heavy (two heaviest) mass eigenstates have been traveling with non-relativistic velocities for billions of years. For a $0.05 \, \mathrm{eV}$ neutrino, this translates to velocities between $\mathcal{O}(10)$ and $\mathcal{O}(10 \, 000)$ km/s, reaching less than ${\sim}$5\% of the speed of light. Therefore, for the time and energy scales we consider in our simulations, it is feasible to describe the trajectories of the massive neutrino states using the nonrelativistic approximation.

The evolution of the cartesian coordinates $x_i$ and momentum $p_i$ of a neutrino is then given by Hamilton's equations
\begin{equation} \label{eq:EOMs}
p_i = a m_\nu x_i', \quad p_i' = -a m_\nu \frac{\partial \Phi(\vec{x}, t)}{\partial x^i}, \quad \mathrm{with}\;\; x_1, x_2, x_3 = x,y,z.
\end{equation}
Here, $a$ is the scale factor, $m_\nu$ is the mass of the neutrino, $\Phi$ is a gravitational potential at position $\vec{x}$ and time $t$. The prime notation denotes the derivative with respect to conformal time. Numerically solving these equations would give the trajectories of a neutrino with one specific mass, and the simulations would therefore have to be re-run, to obtain neutrino trajectories with a different mass.
Previous works have shown how to circumvent this and speed up the computations with two changes: (i) instead of the mass-dependent momentum, solve for the velocity $u_i = p_i/m_\nu$ of the neutrino, effectively replacing $p_i$ with $u_i$ and $m_\nu$ with 1 in eq.~\ref{eq:EOMs} and (ii) treat time as an additional variable to be integrated \cite{Zhang:2017ljh, Mertsch:2019qjv}. The second point ensures that the differential equations become autonomous and still separable, as required by most numerical integration solvers. Choosing this new time variable to be of the form\footnote{For a $\Lambda$CDM cosmology and a negligible radiation energy density for the redshift range considered in our simulations, this time variable takes the form $s(z) = - H_0^{-1} \int_0^z dz \, (1+z)/\sqrt{\Omega_{m} (1+z)^3 + \Omega_{\Lambda}}$, where $\Omega_{m}$ and $\Omega_{\Lambda}$ are the matter and dark energy densities today, respectively.}
\begin{equation}
s(z) = - \int_0^z \frac{dz}{a'}
\end{equation}
fullfills the second point and allows us, together with the first point, to write the equations of motion in the simple form of
\begin{equation}
\frac{dx_i}{ds} = u_i, \quad \frac{du_i}{ds} = -a^2 \frac{\partial \Phi}{\partial x_i}.
\label{eq:EOMs_ui}
\end{equation}
These equations of motion are now solved backwards in time to reconstruct the neutrino trajectories, which represent the paths they took through the gravitational potentials before reaching Earth. We describe the initial conditions of the relic neutrinos and our setup in section~\ref{sec:initialization}. We now run one simulation with a fixed value of the neutrino mass. Then, by appropriately rescaling the output, we can get trajectories of neutrinos with a different mass, without re-running the simulation. We give more details on the procedure and validity of this rescaling method in section~\ref{sec:initialization}. The output of these simulations of most interest to us is the momentum the neutrinos have at the final redshift of the simulation (corresponding to the earliest considered redshift) of $z \approx 4$.\footnote{In~\cite{Tully:2021key} the growth of CNB anisotropies was analyzed and the fluctuations in temperature today was deemed to change by approximately $\pm 0.15 \, \mathrm{K}$ across the sky. Since the deviations were therefore even smaller at the final redshift $z \approx 4$ in our simulations, the assumption of a Fermi-Dirac distribution for the relic neutrinos at this redshift then seems to be reasonable.} The reason for this will become clear in a moment. The number density today is calculated with
\begin{equation}
n_\nu = \frac{g_\nu}{2 \pi^2} \int dp_0 \, p_0^2 \, f_{\mathrm{today}}(p_0),
\label{eq:number_density}
\end{equation}
where $g_\nu$ are the internal degrees of freedom of neutrinos, $p_0$ their momentum today and $f_{\mathrm{today}}(p_0)$ their phase-space distribution of today. We cannot assume a Fermi-Dirac distribution today, as the neutrinos experienced non-linear effects from the gravitational potentials during their propagation, which we explore in more detail in section~\ref{sec:results_phase_space}. However, we can use Liouville's theorem --- the fact that phase-space density is conserved for trajectories obeying the equations of motion --- to get the phase-space distribution value today with
\begin{equation}
f_{\mathrm{today}}(p_0(z_0), z_0) = f_{\mathrm{FD}}(p_f(z_f), z_f).
\label{eq:liouville}
\end{equation}
Here, the subscripts $0$ and $f$ on the momentum $p$ and redshift $z$ denote the value at the start and end of our calculations, respectively. The penultimate step is to convert the simulated velocity at the final redshift, $u_f$, back to momentum for a specific target neutrino mass simply via $p_f = u_f \cdot m_{\mathrm{target}}$, then we use Eqs.~\ref{eq:number_density} and~\ref{eq:liouville} to obtain the local number density for this neutrino mass today. Finally we compute the clustering factor $f_i \equiv n_{\nu}/n_{\nu,0}$, i.e. the ratio of the number density of our simulation to the cosmological one for one family of neutrinos plus anti-neutrinos.\footnote{This means we took $g_\nu=2$ in eq.~\ref{eq:number_density} throughout this work.} This predicted number density from $\Lambda$CDM cosmology, $n_{\nu,0}$, is obtained by computing the integral in eq.~\ref{eq:number_density} with the Fermi-Dirac distribution, which results in 
\begin{equation}
n_{\nu,0} = \frac{3\zeta(3)}{2\pi^2}T_{\nu,0}^3 \approx 112 \, \mathrm{cm}^{-3}.
\label{eq:number_density_norm}
\end{equation}

\section{Methodology}
\label{sec:methods}

The primary objective of this study is to investigate the effects that manifest when transitioning to a comprehensive N-body treatment of the structure in the local Universe. In previous studies it was feasable to calculate gravitational forces felt by the neutrinos at each time step in the simulation, where the equations for the gravitational potentials obeyed some symmetry, e.g., spherical or cylindrical. Instead of smooth potentials, however, we are dealing with individual particles distributed in a non-uniform manner as an input. The individual neutrinos are simulated in the resulting complex gravitational environment.

With the evidence-supported assumption that the energy density of DM dominates the total gravitational potential, the DM halos will not feel the impact of the (mostly) non-relativistic neutrinos in a significant manner during the timescales we consider. The gravitational interaction of the relic neutrinos with each other is also negligible. This allows us to adopt the ``N-1-body'' approach first laid out in \cite{Ringwald:2004np}, allowing us to trace the neutrinos individually and benefit from parallelized simulations.

In the following sections, 
we describe the CDM-only cosmological simulations of~\cite{Correa:2022dey}, the calculations of the gravitational forces, the initial conditions of the neutrinos and what type of simulations we ran for what purpose.

\subsection{Cosmological Simulations}
\label{sec:TangoSIDM}

In this work we make use of a cosmological DM-only simulation from the {\sc TangoSIDM} project. It comprises DM-only and hydrodynamical cosmological volumes of 25 Mpc on a side~\cite{Correa:2022dey,Correainprep}. The simulations have been produced using the SWIFT\footnote{https://swiftsim.com} code~\cite{Schaller16,Schaller18,Schaller23} that include various hydrodynamics and gravity schemes. While the goal of the project is to investigate the impact of self-interacting DM on the formation of cosmic structures, the analysis in this work focuses on the collisionless CDM-only simulations of (25 Mpc)$^{3}$ that follows the evolution of 752$^{3}$ DM particles, reaching a spatial resolution of 650 pc and a mass resolution of $1.44\times 10^{6}~\rm{M}_{\odot}$.\footnote{The comoving softening is 1.66 kpc at early times and freezes at a maximum physical value of 650 pc at redshift $z = 2.8$.} The adopted cosmological parameters relevant for this work are the matter and dark energy densities of $\Omega_{\rm{m}}=0.307$ and $\Omega_{\Lambda}=0.693$, respectively, as well as the dimensionless Hubble constant of $h=0.6777$. For this work, we extract CDM-only halos from the 25 Mpc box, which we henceforth refer to as box halos (or box halo sample) and cosmological box, respectively. The output of these simulations are stored only at specific redshifts, in so-called snapshots, and the only relevant part for us are the data of the coordinates for all DM particles of the halos of these snapshots. The resulting evolution of the DM particles is used to construct the gravitational grid for these specific redshifts to calculate the neutrino trajectories. By using 25 logarithmically spaced snapshots between redshifts 0 and 4, we were able to reliably produce results with our methodology, which were comparable to the results obtained with the analytical approach.

The simulation suite contains a full set of halo catalogues and merger trees that were generated using the VELOCIraptor halo finder \cite{Elahi11,Elahi19,Canas19}. VELOCIraptor uses a 3D-friends of friends (FOF) algorithm \cite{Davis85} to identify field halos, and subsequently applies a 6D-FOF algorithm to separate virialised structures and identify subhalos of the parent halos \cite{Elahi19}. To link halos through time, the halo merger tree code TreeFrog \cite{Elahi19b} was used. Throughout this work, the virial halo mass $M_{200}$, is defined as all matter within its virial radius $R_{200}$, for which the mean internal density is 200 times the critical density of the Universe. In each FOF halo, the ``central'' halo is the halo closest to the center (the minimum of the potential), which is nearly always the most massive. The remaining halos within the FOF halo are its satellites, also called subhalos. The halo catalogues provide virial masses and radii for subhalos, as well as for the main halos, and the concentration parameter $c_{200}$, defined as the ratio between $R_{200}$ and the scale radius, $R_{s}$, the latter being the radius at which the logarithmic density slope is -2. The particle mass resolution of the simulations is sufficient to resolve (sub)halos down to ${\sim}10^{7}~\rm{M}_{\odot}$ with $10$ particles.

\subsection{Gravity Calculations}
\label{sec:precalculation}

A widely utilized approach for large-scale numerical simulations, which we also employ in the current work, involves partitioning the simulation space into uniform subdivisions in three-dimensional space, specifically into cubic segments. Hereafter, these segments will be referred to as cells. Instead of uniformly sized cells throughout the simulation space, we adapt the cell size to the DM content with the following three-step algorithm.
\begin{enumerate}
    \item {\it First Cell.} A single cell is created to encompass all the chosen structure(s) the cosmological simulation box.

    \item {\it Cell Division.} The cell undergoes an iterative octree division process in which each cell is partitioned into eight smaller, equally sized cells if the number of particles contained surpasses a predetermined threshold.\footnote{We made this threshold dynamical, such that it depends on the distance of the cell to the host halo center. The farther away a clump of DM, the bigger the smallest cell is allowed to be. This reduces computation time without the loss of crucial information, since the neutrinos reach these far away parts later in the simulation (i.e. earlier in time), where the (dark) matter distribution is more homogeneous.}

    \item {\it Cell Gravity.} The force that a particle would experience at the center of each cell is computed. The gravitational strength is then treated uniformly throughout the entire cell. 
\end{enumerate}
The minimal cell size is determined by the upper bound of allowed DM particles in each cell. A trade-off exists in relation to cell size: smaller cells yield a more precise gravitational grid, but at the expense of increased computational time.\footnote{We chose the threshold to be 1000 DM particles.} For the final step, the gravitational force for a cell is split up into a short-range and long-range force, denoted with subscript sr and lr in the following equations, respectively. The short-range force results from the superposition of forces exerted by all DM particles present in a cell, whereas the long-range force arises from the superposition of forces due to all DM contained in all other cells in the grid. For the remainder of this section, we refer to the cell for which gravity is computed as cell C, and label any other cell with J, with particles therein designated by the subscript $j$.

The short-range potential at the central position $\vec{r} = (x,y,z)$ of cell C is given by
\begin{equation}
\Phi_{\mathrm{sr}}(\vec{r}) = -G \sum_i \frac{m_i}{\sqrt{\norm{\vec{r} - \vec{r}_i}^2 + \varepsilon^2}},
\end{equation}
where the sum goes over all the DM particles with individual mass $m_i$ and distance $\vec{r}_i$ contained in cell C. In the cosmological simulation box we are using, all the DM particles have the same mass ($m_{\mathrm{DM}}$ hereafter). The small $\varepsilon$ offset corresponds to the gravitational softening\footnote{For an in-depth discussion on gravitational softening, see, e.g., \cite{Springel:2020plp}.} used in the {\sc TangoSIDM} simulation suite. Denoting the cartesian coordinates of the center of cell C (i.e. the components of $\vec{r}$) with $x_a$, where $a = 1,2,3$ and $x_1, x_2, x_3 = x, y, z$, the derivative at the coordinate $x_a$ is then
\begin{equation}
\frac{\partial \Phi_{\mathrm{sr}}}{\partial x_a} = G \, m_{\mathrm{DM}} \sum_i \frac{x_a - x_{a,i}}{(\norm{\vec{r} - \vec{r}_i}^2 + \varepsilon^2)^{3/2}}.
\end{equation}
The sum goes over all DM particles with coordinates $x_{a,i}$ present in cell C. 

The long-range force is computed via a multipole expansion of the Newtonian potential. Keeping the first 3 moments, the potential caused by the DM content in a single cell J is then
\begin{equation}
\Phi_{\mathrm{lr}}(\vec{r}) = - G \, m_{\mathrm{DM}} 
    \left( 
    \frac{1}{\norm{\vec{r} - \vec{r}_J}} 
    + \sum_a \frac{D_a x_a}{\norm{\vec{r} - \vec{r}_J}^3} 
    + \sum_{a,b} \frac{Q_{ab} x_a x_b}{2 \norm{\vec{r} - \vec{r}_J}^5}
    \right).
\label{eq:lr_orig}
\end{equation}
The first term corresponds to the monopole, the second term contains the dipole moment, $D_a = \sum_{j \in J} x_{j,a}$, and the last term contains the quadrupole moment as a $3\times 3$ matrix with elements $Q_{ab} = \sum_{j \in J} (3 x_{a,j} x_{b,j} - \norm{\vec{r}_j}^2 \delta_{ab})$. The sums go over all DM particles contained in cell J. We take all coordinates and distances with respect to the center of mass of cell J, $\vec{r}_J$, which makes the dipole moment vanish. In this frame of reference, the derivative at position $x_a$\footnote{In eq.~\ref{eq:lr_orig}, the coordinate $x_a$ is centered on the center of the cell J, whereas now in eq.~\ref{eq:lr_final} it is centered on the center of mass of cell J, $\vec{r}_J$.} is
\begin{equation}
\frac{\partial \Phi_{\mathrm{lr}}}{\partial x^a} = G \, m_{\mathrm{DM}} 
    \left(
    \frac{x_a}{\norm{\vec{r} - \vec{r}_J}^3}
    - \frac{1}{\norm{\vec{r} - \vec{r}_J}^5} \sum_{a,b} Q_{ab} x_b
    + \frac{5 x_a}{2 \norm{\vec{r} - \vec{r}_J}^7} \sum_{a,b} Q_{ab} x_a x_b
    \right).
\label{eq:lr_final}
\end{equation}

In our simulations we have thousands of cells and millions of individual DM particles. We do not compute the quadrupole moment for all J cells. It is only computed for cells whose center of mass coordinates have a distance from cell C below a critical distance. Our ad hoc criterium for this critical distance is $r_\mathrm{crit} = 1.5  L_C  G_C^{0.6}$, where $L_C$ and $G_C$ are the length and the generation of cell C, respectively. Generation 0 would be the initial cell placed over the total DM content. A cell division step results in the ``parent'' cell being divided into 8 ``child'' cells, each of which carries a generation index increased by 1. This ad hoc formulation was chosen by trial and error, such that the qudrupole moments are computed only for the immediate neighbouring cells for the outer-region cells. For the inner-region cells, however, where the DM distribution can be more dense and complex, there is a larger number of cells worth computing the quadrupole moment for. In this way, the monopole is computed for all cells, whereas the quadrupole moment is computed where most impactful and necessary. The total long-range force in cell C is then the superposition of the forces due to all other cells in the grid.

The fastest moving neutrinos can reach coordinates outside the grid. For such cases, we treat the potential at those coordinates as one caused by a point mass at the center of mass coordinates ($x_\mathrm{c.o.m.}$) of all DM particles in the grid. The point mass here is the total mass $m_{\mathrm{tot}} = N_{\mathrm{tot}} \, m_{\mathrm{DM}}$, with $N_{\mathrm{tot}}$ the total amount of selected DM particles. The gradient at position $x_a$ is then simply
\begin{equation}
\frac{\partial \Phi}{\partial x^a} = G \, m_{\mathrm{tot}} \frac{x_a - x_\mathrm{c.o.m.}}{\norm{\vec{r} - \vec{r}_\mathrm{c.o.m.}}^3}.
\end{equation}

\subsection{Initial Conditions}
\label{sec:initialization}

One of the goals of this work is to precisely determine the local number density of relic neutrinos at Earth. Earth's distance from the Galactic center (GC) has recently been determined to be ${\sim}8.178$ kpc~\cite{2019A&A...625L..10G}. We therefore inject the relic neutrinos at a similar distance from the center of potential of the considered DM halos in our simulations. For spherically symmetric simulations, where the potentials are described analytically, the starting distance can be exactly this Earth-GC distance. 
For our dicrete simulation space, however, the centers of the cells are at fixed locations. Therefore, the neutrinos in our simulations start in a cell, which has a distance from the halo center close to this Earth-GC distance, but not exactly. Depending on the resolution of our spatial grid, which depends on the composition of the halo, the starting distance can be off by up to ${\pm}1$ kpc. However, we show in section~\ref{sec:results_number_density_band}, that this variance in initial distance has a negligible impact on our results.

We took logarithmically spaced values for the magnitude of the momentum in the range from 0.01 to 400 times the CNB temperature. These limits are based on the previous works on neutrino clustering \cite{Ringwald:2004np, deSalas:2017wtt, Zhang:2017ljh, Mertsch:2019qjv}, but have been adjusted empirically. As a reminder, the main advantage is that we only need to run the simulation once with a fixed neutrino mass. The results can then be rescaled to other neutrino masses, as briefly mentioned in section~\ref{sec:theory}. However, the neutrino mass cannot be arbitrarily low. 
Since we have a fixed momentum range, the limits of the corresponding velocity interval increase as you decrease the neutrino mass. The validity and accuracy of the equations of motion in this ever-more relativistic regime should then be questioned. We therefore opted to use a neutrino mass of $0.3$ eV\footnote{Although the $\Lambda$CDM constraint on the sum of neutrino masses of $\sum m_\nu \lesssim 120 \, \mathrm{meV}$~\cite{Planck:2018vyg} would exclude our chosen upper limit, such constraints are model-dependent. The constraints can vary and relax substantially with different cosmologies and datasets (see e.g. \cite{Escudero:2020ped}). Our maximum neutrino mass is also still allowed by the current KATRIN limit of $m_\nu \lesssim 0.8$ eV~\cite{KATRIN:2021uub}.} to run the simulations with, together with the upper limit of $400 T_{\nu,0}$ for their momentum. This means that the maximum velocity in our simulations is ${\sim}22\%$ the speed of light.\footnote{Due to the nature of the Fermi-Dirac distribution, this means that only the neutrinos in the very upper end of the distribution will be moving with these velocities. They do not make up the bulk of our simulated relic neutrinos, and hence, the classical treatment is valid for the majority of our simulation content.} We chose this upper limit to guarantee sufficient coverage of the momentum distribution for the lightest neutrino mass of $0.01$ eV we considered. However, the rescaling is only effective for this range of neutrino masses. For neutrino masses outside of this range, the coverage of their momentum distribution becomes inadequate, and the integrand in eq.~(\ref{eq:number_density}) is truncated prematurely, leading to an underestimation of the number density.

The neutrinos are traced back until redshift $z \approx 4$. There is one crucial difference between the analytical method and our numerical one; the frequency of updates of the gravitational grid during the simulation. With the use of analytical functions for the halo parameters it is possible to update the gravitational forces at each time step. However, since we have a fixed number of snapshots from the cosmological box, we can only update the gravitational potential at redshifts corresponding to these snapshots. With 100 redshift steps and 25 snapshots we were able to reliably reproduce previous results. To solve the equations of motion numerically, we made use of the differential equation solver \verb|solve_ivp| from the \texttt{scipy} library. For our purposes, the Runke-Kutta method of order 2(3) was sufficient and the use of higher orders showed no significant impact on our results.

\subsection{Analysis Strategy}
\label{sec:types_sims}

Our analysis is split into two parts, for which we developed 2 types of simulations for the reconstruction of the neutrino trajectories, following the methodology described in section~\ref{sec:methods}. These types use the same gravitational grid and only differ in the initial conditions of the neutrinos and the saved outputs. The first type is dedicated towards analyzing the overall local number density, whereas the second type is used to generate and inspect CNB all-sky anisotropy maps.
 
The goal of the first type of simulations is to investigate how the uncertainties for the parameters describing the MW (e.g. the virial mass) and the DM distribution of its halo impact the local relic neutrino number density. Various works have estimated the MW mass using the latest Gaia data \cite{Karukes:2019jwa,2019A&A...621A..56P,2019ApJ...875..159E,2018A&A...619A.103F}, and have found values withing the range of $M_{200} = (0.6\mbox{--}2.0) \times 10^{12} \, \mathrm{M_\odot}$. We were able to find 27 viable halos (in the considered cosmological box described in section~\ref{sec:TangoSIDM}) with a virial mass in this range, which have very different DM distributions.
The azimuthal and polar angles for the initial directions of the neutrinos were split into 20 linearly spaced steps, ranging from 0 to $2\pi$ and 0 to $\pi$, respectively. The momentum was spaced logarithmically in 200 steps between our limits of $0.01 T_{\nu,0}$ and $400 T_{\nu,0}$, resulting in 80 000 neutrinos with unique combinations of direction and momentum for this type of simulation. Beyond this amount we found negligible improvement for our setup, whereas going below could result in irregular curves for some halos. Importantly, in the final integration to obtain the number density, all directional information is lost and we obtain only a single value for the local relic neutrino number density.

The purpose for the second simulation type is to examine the relation between the characteristics of the DM distributions of the halos with the anisotropies of the CNB. To explore this, instead of evenly spaced azimuthal and polar angles, we used the \texttt{healpy} library~\cite{Gorski:2004by} to get the initial angles of the pixels of a healpix all-sky map. We simulated 10 000 neutrinos for each pixel. We noticed that this amount saturated the integral and going below this value sometimes resulted in underestimations of the number densities. For our resolution of Nside=8 (768 pixels), this results in a total of 7.68 million neutrinos for this type of simulation.

All calculations were performed on a 128 CPU core cluster node with 224 GB memory. The run time for the precalculations, i.e. constructing the gravitational grid, could take up to ${\sim}1$ hour for one halo, whereas simulating the neutrinos could take up to ${\sim}3$ hours, depending on the type of analysis.

\section{Results}
\label{sec:results}
The results of this work are divided into two parts. First, we address the local number density of relic neutrinos in the vicinity of Earth. This part is also used for validating our methodology by recreating results of previous works and comparing them to the predictions of our simulations. Second, we document our findings regarding the anisotropies which arise due to the non-linear dynamics of our simulations by specifically calculating the expected number densities for different directions in the sky.

\subsection{Local Number Density}
\label{sec:results_number_density_band}

\begin{figure}[ht!]
    \centering
    \includegraphics[width=.6\textwidth]{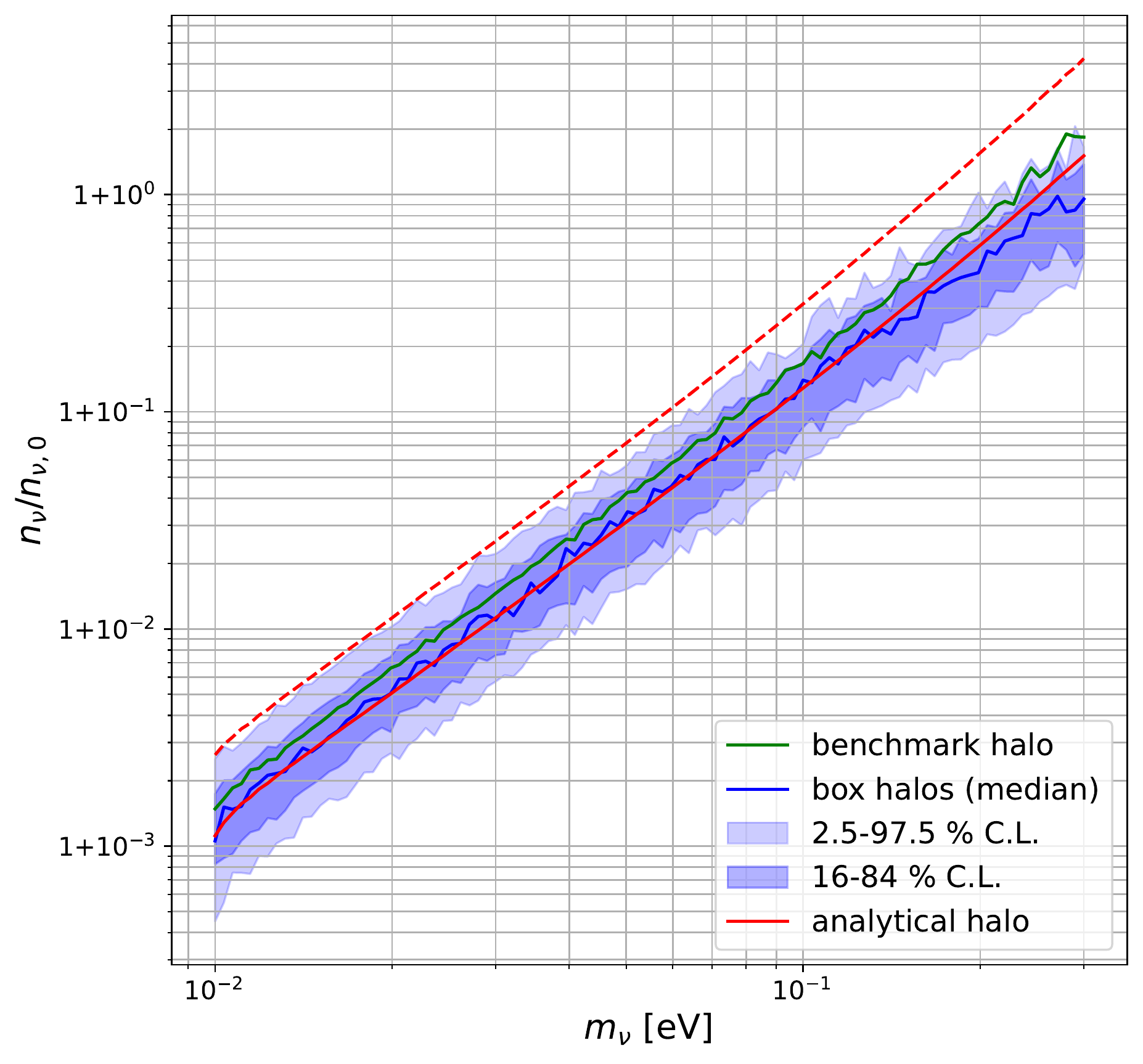}
    \caption{Relic neutrino clustering factors as a function of the neutrino masses considered in this work. The red curves are produced with the analytical approach for the gravitational potentials from \cite{Mertsch:2019qjv}. More specifically, the red dashed curve corresponds to their results (using their parameters for the MW type halo), whereas using parameters from our selected halos from the cosmological simulation box produces the red solid curve. Results from our approach are displayed in green and blue (see text for more details).}
    \label{fig:overdensity_band}
\end{figure}

We first performed a simulation with the purpose of comparing our approach to the one used previously in~\cite{Mertsch:2019qjv}. For this we created a spherically symmetric halo made up of individual particles, based on a Navarro-Frenk-White (NFW) density distribution~\cite{Navarro:1995iw}. We describe the construction process in appendix~\ref{app:NFW_halo}. The resulting number densities for different neutrino masses from this simulation are shown in figure~\ref{fig:overdensity_band} as the green solid curve. This is to be compared to our recreation of the results of~\cite{Mertsch:2019qjv} depicted by the red solid curve. For a fair comparison, we used the median for the virial mass, virial radius and concentration of our cosmological box halo sample to produce both these curves, such that they only differ in methodologies. Our simulation framework seems to be validated by the fact that there is a good level of agreement between these curves from these two different methodologies.

As previously stated, the largest cosmological box available to us was the 25 Mpc box and we were able to find 27 viable DM halos, whose virial mass was within the required range of $M_{200} = (0.6$--$2.0) \times 10^{12} \, \mathrm{M_\odot}$. The values of the virial radius and concentration were in the range of $R_{200} \approx 178$--$257 \, \mathrm{kpc}$ and $c_{200} \approx 6$--$15.5$, respectively.  The value of the initial distance varies depending on the halo but was within the range of $d_{\mathrm{init}} \approx 7.2$--$9.1  \, \mathrm{kpc}$. The blue solid curve in figure~\ref{fig:overdensity_band} corresponds to the median values for the number densities obtained with the selected halos, while the dark and light shaded blue regions show the 68\% and 95\% containment regions, respectively. Overall, the two methodologies produce curves with similar normalisation and slope, when the halo parameters are similar. As expected, we see a spread from our numerical simulations resulting from the different characteristics of the halos. The clustering factors predicted by our numerical simulations can drop to negligible values of ${\sim}1.0005$ for $0.01$ eV neutrino masses and reach up to ${\sim}3$ for $0.3$ eV neutrino masses. When using the same MW parameters as in \cite{Mertsch:2019qjv}, which are close to our considered upper limits\footnote{Specifically for \cite{Mertsch:2019qjv}: $M_\mathrm{vir} = 2.03 \times 10^{12} \, \mathrm{M_\odot}$, $R_\mathrm{vir} = 333.5 \, \mathrm{kpc}$ and $c_\mathrm{vir} \approx 16.8$.}, and using their methodology, we recover their results as depicted by the red dashed curve, which can differ up to an order of magnitude to our results. As we confirmed earlier, the two methodologies produce similar results when the halo parameters are similar. Therefore, the only reason for this discrepancy are the different halo parameters.

For higher masses, the blue band begins to slightly dip below the values obtained via the other methods. We comment and speculate on the reasons for this in section~\ref{sec:results_phase_space}, when we investigate the phase space the relic neutrinos occupy today.

\begin{figure}[ht!]
    \centering
    \includegraphics[width=.9\textwidth]{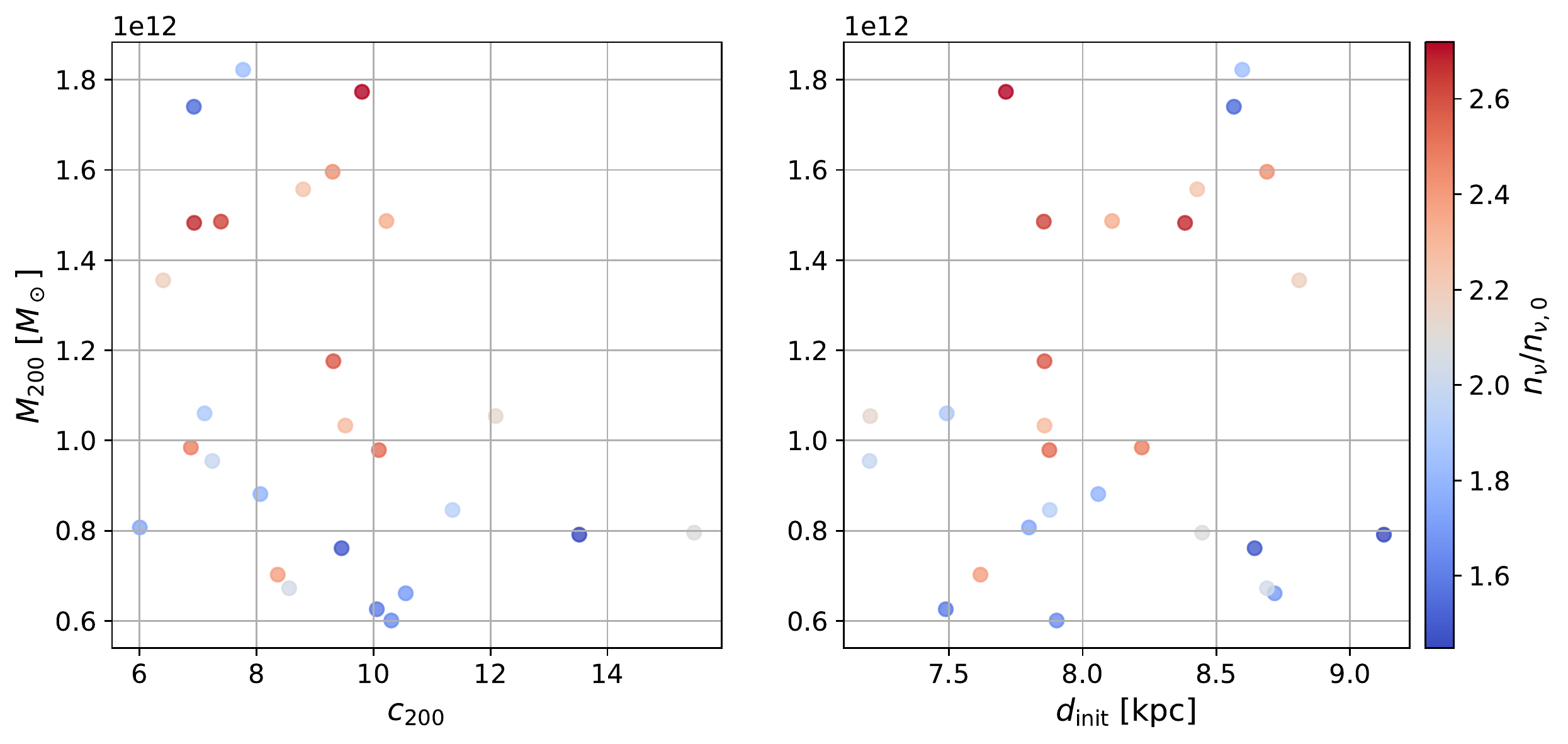}
    \caption{The local relic neutrino overdensities in the $M_{200}$--$c_{200}$ and $M_{200}$--$d_\mathrm{init}$ parameter planes are shown in the left and right panel, respectively, for $0.3$ eV neutrino masses. In general, the clustering factors are correlated with the halo mass, whereas they seem to be rather independent of concentration and initial distance of the starting cell from the center of the halo, i.e. Earth's position in the simulation.}
    \label{fig:2D_params_box_halos_0.3eV}
\end{figure}

As the virial mass, concentration and initial neutrino distance vary accross the halos, we show the clustering factors for $0.3$ eV neutrinos in two parameter planes in figure \ref{fig:2D_params_box_halos_0.3eV}. On average the clustering factors increase for more massive halos, rather independent on the concentration or initial distance. The same trends are found for all neutrino masses.

\subsection{Anisotropies}
\label{sec:results_anisotropies}

We now direct our attention to the anisotropies of the CNB. Interestingly, the results vastly differ across methodologies, and there seem to be emerging features when transitioning to our numerical methods. For this section, we use the same halo sample as in section~\ref{sec:results_number_density_band}, where we looked at the overall local number densities. We adopted the common choice to display anisotropies on an all-sky map using the \texttt{healpy} library~\cite{Gorski:2004by}. All the maps in this section are displayed with the mollview projection and have $\mathrm{Nside}=8$ for a total of $\mathrm{N_{pix}}=768$ (heal)pixels, resulting in an approximate resolution of $7.33^\circ$ degree for one pixel. The resulting number densities for each pixel, $n_{\nu,\mathrm{pix}}$, are normalized to the standard cosmological value distributed among all pixels, i.e., the normalization for one pixel is $n_{\nu,\mathrm{pix},0} = n_{\nu,0} / \mathrm{N_{pix}}$. We used galactic longitudes and latitudes, $(l,b)$, for the coordinates in the sky and have rotated the maps where necessary, such that the center of potential of the halos are at the center of the maps with $(l,b) = (0^\circ,0^\circ)$. We refer to the opposite point at $(l,b) = (180^\circ,0^\circ)$ as the anti-center. Since the rotation of healpix maps sometimes causes certain pixels to contain no values, we have interpolated these pixels with the mean of their nearest neighbours.

\begin{figure}[ht!]
    \centering
    \includegraphics[width=1.\textwidth]{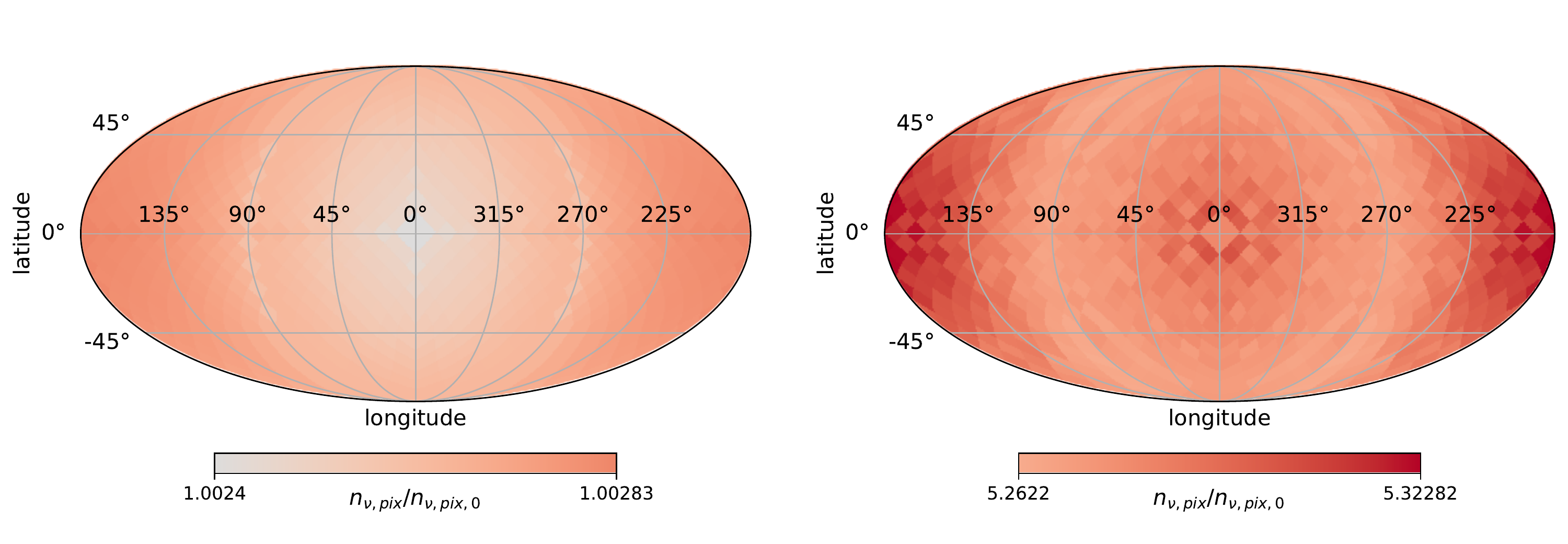}
    \caption{All-sky anisotropy map in galactic coordinates from the analytical approach. The left and right panels show the clustering factors for $0.01$ and $0.3$ eV neutrinos, respectively.}
    \label{fig:All_sky_maps_analytical}
\end{figure}

First, we show the results obtained with the analytical approach in figure~\ref{fig:All_sky_maps_analytical}. We see that there are only enhancements to the number density for any pixel and for any of the considered neutrino masses. We show only the endpoints of the considered neutrino mass spectrum to highlight the difference in the patterns and numerical values for the pixels. The overall trends are: (i) the higher the mass the higher the number densities in any pixel and (ii) the higher the mass the higher the difference between the minimum and maximum number density. However, for the second point, the percentege difference is at most of order $\mathcal{O}(1)$ for $0.3$ eV neutrinos. For $0.01$ eV neutrinos the number densities are effectively isotropic with negligible enhancements in any direction.
Due to the smooth evolution of the halo and the lack of DM substructure, the gravitational focussing seems to create undisrupted geodesics for the neutrinos, as well as altering the path of some neutrinos that would have missed Earth otherwise, resulting in overdensities everywhere in the sky for all neutrino masses. The values in the direction of the center are lower than those in the direction of the anti-center, since the loss of momentum when rising out of the potential well decreases their probability of reaching Earth.

\begin{figure}[ht!]
    \centering
    \includegraphics[width=1.\textwidth]{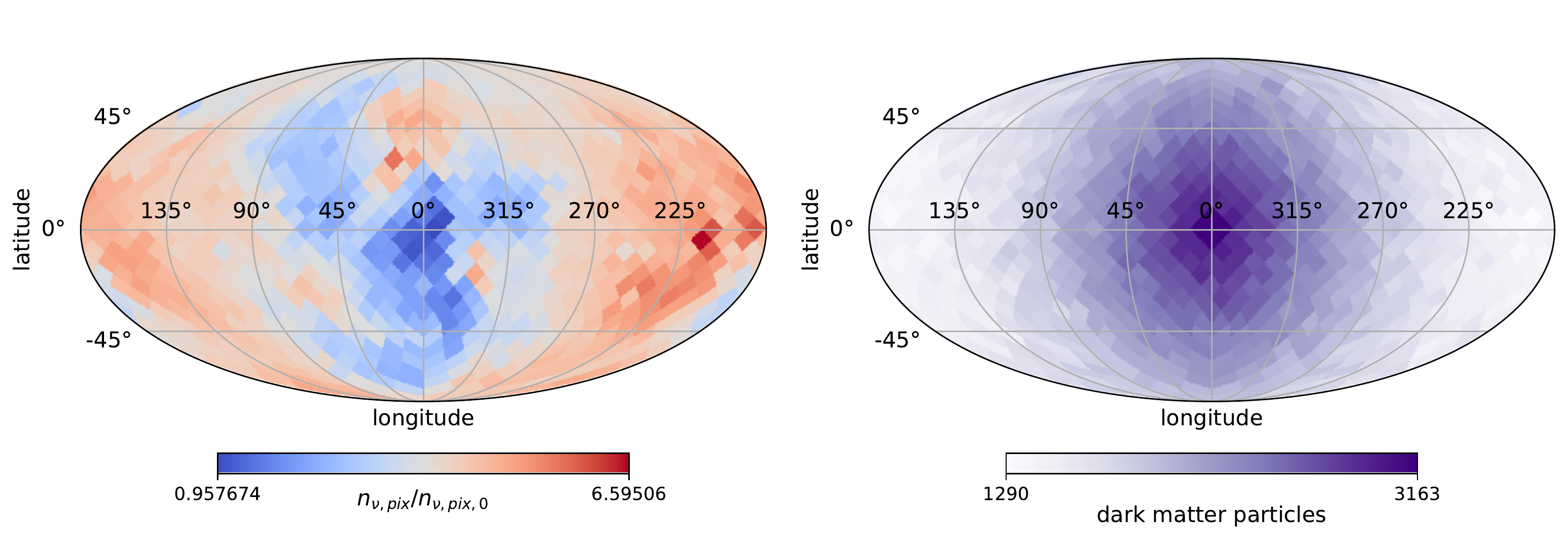}
    \caption{All-sky anisotropy map in galactic coordinates from our benchmark simulation. The left panel shows the clustering factors for $0.3$ eV neutrinos. The right panel shows the projected DM content in the directions the pixels represent.}
    \label{fig:All_sky_maps_benchmark_halo}
\end{figure}

There are crucial differences predicted by our numerical simulations. The most striking difference is the appearance of both over- \emph{and} underdense regions. This is first demonstrated with our benchmark simulation. In figure~\ref{fig:All_sky_maps_benchmark_halo} we see an overall anticorrelation of the pixel number densities (left panel) with the amount of DM in that direction (right panel), resulting in an underdense region towards the center. 
From figure~\ref{fig:overdensity_band} in the previous section we know that the overall local number density between the methodologies does not change significantly. The change comes in the form of a re-distribution of the number densities across the sky. While the enhancement was more prominent at the anti-center, dropping to lower values at $90^\circ$ angles, and rising again slightly towards the center with the analytical approach, the benchmark simulation shows a somewhat smooth increase of number densities from the center to the anti-center. Finally, the difference between the lowest and highest value is now much larger than with the analytical approach, reaching a factor of ${\sim}7$ for the benchmark simulation.

\begin{figure}[ht!]
    \centering
    \includegraphics[width=1.\textwidth]{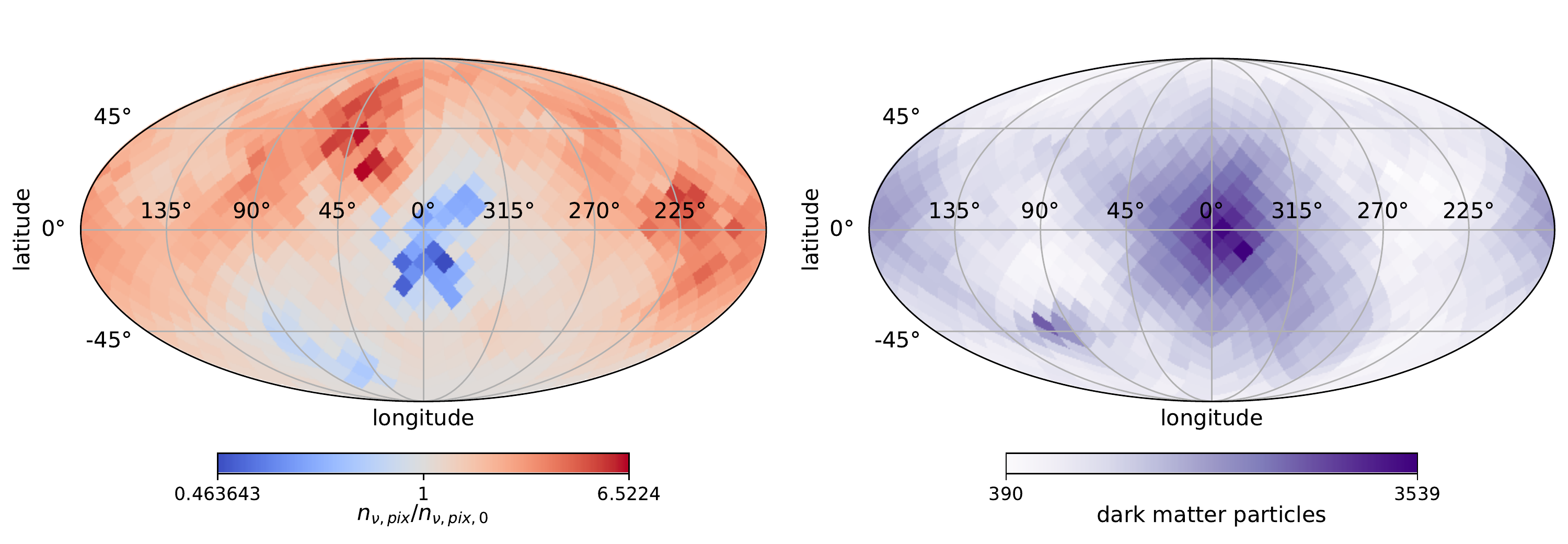}
    \caption{All-sky anisotropy map for a halo from the cosmological box, where the number densities are somewhat spatially anti-correlated with the DM content.}
    \label{fig:All_sky_maps_halo13}
\end{figure}

\begin{figure}[ht!]
    \centering
    \includegraphics[width=1.\textwidth]{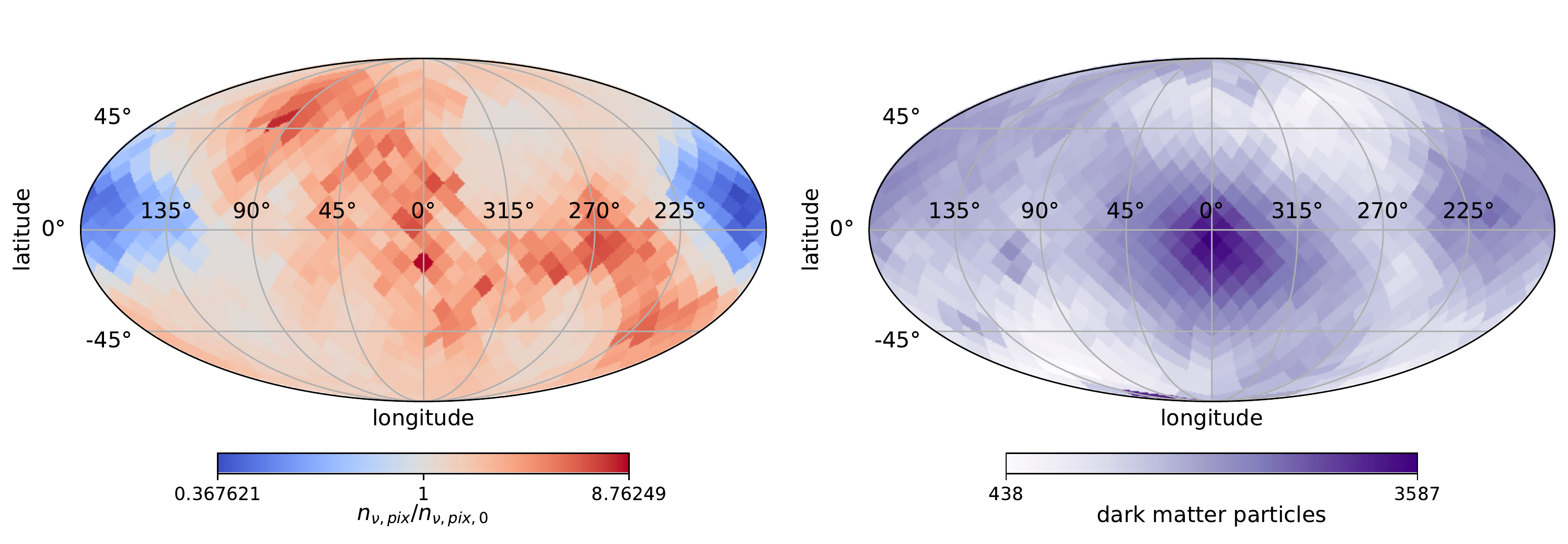}
    \caption{All-sky anisotropy map for a halo from the cosmological box, where the number densities are somewhat spatially correlated with the DM content.}
    \label{fig:All_sky_maps_halo14}
\end{figure}

Regarding the all-sky maps of our box halo sample, the values and patterns vary, with some trends being the same as before. The common trends predicted by these simulations, regardless of which halo was used, are (i) the existence of over- and underdense regions, (ii) a large range of values for the individual pixels, differing with factors up to ${\sim}5$ and (iii) the re-distribution of number densities across the sky, sometimes displaying spatial (anti)correlations with the DM content along the line of sight. We show two halos in particular to showcase some features arising from this simulation approach in figure~\ref{fig:All_sky_maps_halo13} and figure~\ref{fig:All_sky_maps_halo14}. 

\begin{figure}[ht!]
    \centering
    \includegraphics[width=1.\textwidth]{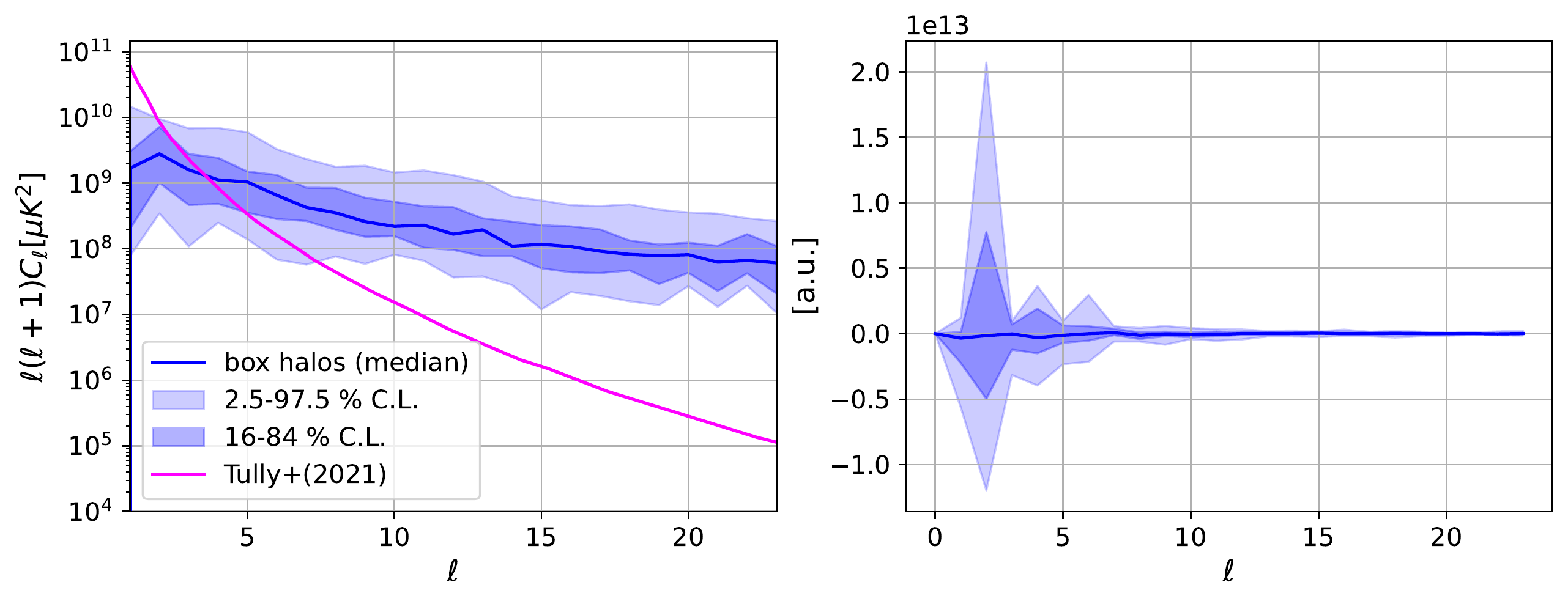}
    \caption{Left: Angular power spectra of our CNB anisotropy maps for $0.1$ eV neutrinos of the box halo sample. The blue solid band are the median values with the 68\% and 95\% coverage band as the dark and light blue regions, respectively. For comparison we show the power spectrum for the same neutrino mass obtained in~\cite{Tully:2021key} in magenta. Right: Cross-power spectra in arbitrary units of our anisotropy maps with the corresponding projected DM content map. The $C_\ell$ were computed with the \texttt{healpy} library.}
    \label{fig:all_power_spectra}
\end{figure}

To draw parallels to previous works on CNB anisotropies, we compute the angular power spectra of our number density anisotropy maps. Specifically we compare our spectra to the recent results of~\cite{Tully:2021key}, for which we first converted the number densities to brightness temperatures via eq.~(\ref{eq:number_density_norm}) to match their units. The authors of this study have computed the temperature fluctuations caused by the matter inhomogeneities, as predicted by the primordial power spectrum, which the relic neutrinos encountered during their travel. We compare our results in figure~\ref{fig:all_power_spectra} in the left panel.
Since we do not include the large-scale structure beyond the virial radius, the power on large angular scales at dipole and quadrupole for our halos is 1--2 orders of magnitudes lower than in~\cite{Tully:2021key}. On smaller scales beyond octopole, however, the power due to the local clustering effect is much larger, by many orders of magnitude. The right panel of figure~\ref{fig:all_power_spectra} shows the cross-spectra between the anisotropy maps and the corresponding projected DM maps. There is no preferred correlation either way, but there do exist halo configurations resulting in anti-correlations and correlations as shown previously in figure~\ref{fig:All_sky_maps_halo13} and figure~\ref{fig:All_sky_maps_halo14}, respectively.

\subsection{Phase-space Distribution}
\label{sec:results_phase_space}

\begin{figure}[ht!]
    \centering
    \includegraphics[width=.8\textwidth]{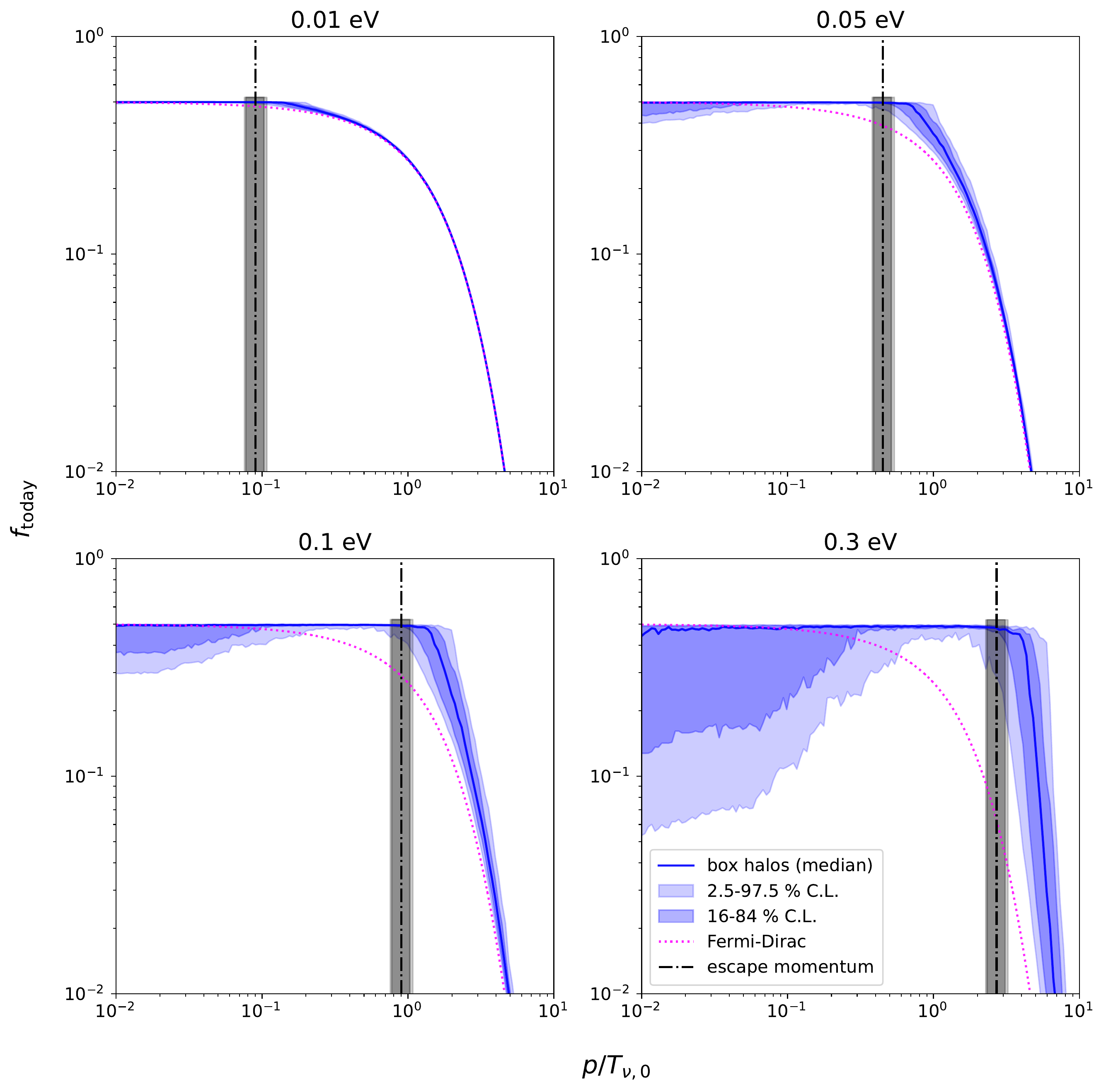}
    \caption{Momentum distributions of today for 4 different neutrino masses. The blue solid line represents the median values of $f_\mathrm{today}$ for our box halo sample, with the dark and light shaded blue regions representing the 68\% and 95\% containment regions, respectively. The (unaltered) Fermi-Dirac distribution is depicted by the dotted magenta line. The black dash-dotted line and the grey shaded area represent different escape momenta (see text for more details).}
    \label{fig:phase_space_numerical_log}
\end{figure}

Relic neutrinos decouple relativistically from the early Universe plasma, and are believed to follow a Fermi-Dirac phase-space distribution at that time, with very small corrections.\footnote{See, however, ref.~\cite{Alvey:2021sji} on why different initial neutrino distribution functions are also possible.} As relic neutrinos enter their non-relativistic phase, they experience the effects of gravitational potentials more strongly, which can change their phase-space distribution. We show the momentum distributions of today, $f_\mathrm{today}$, for 4 different neutrino masses of $(0.01, 0.05, 0.1, 0.3)$ eV in figure~\ref{fig:phase_space_numerical_log}, which were possible to compute with eq.~(\ref{eq:liouville}). 
The neutrinos in our simulations are initialized with multiple directions for each momentum. Each direction for the same momentum has thus a different value of $f_\mathrm{today}$. Lower values correspond to fewer (on average) neutrinos coming from that direction. We only display the highest value for each momentum for visual clarity. Since the overall number density integrates over all momenta, the set displayed in figure~\ref{fig:phase_space_numerical_log} gives the largest contribution. Our conclusions based on this figure can therefore be generalized.

The blue solid curve shows the median values of $f_\mathrm{today}$ of the box halo sample and the dark and light shaded blue regions contain 68\% and 95\% of the values, respectively. We show the unaltered Fermi-Dirac distribution as the magenta dotted curve. There are two ways in which the momentum distribution predicted by our simulations deviates from a Fermi-Dirac one: (i) more higher momentum states are filled (up to a certain threshold) and (ii) more lower momentum states get depleted. Both effects become more pronounced as the neutrino mass increases. 

The first effect is caused by the escape condition the neutrinos have to fullfill to not get captured by the gravitational potential well at the location of Earth, $\Phi(x_\oplus, y_\oplus, z_\oplus)$. Up until an ``escape momentum'' the momentum states seem to be fully occupied, since they are more likely to be (fully) captured. Above this threshold, the momentum states are less likely to be occupied as the neutrinos can leave the galaxy, and $f_\mathrm{today}$ becomes Fermi-Dirac-like again. The escape momentum can be approximated by $p_\mathrm{esc} = \sqrt{2 |\Phi(r_\oplus)|} m_\nu$.\footnote{This formula assumes that the potential is spherically symmetric, which is only approximately true for some of the box halos. We see that the momentum states are mostly occupied until these escape momentum values and quickly deplete afterwards. However, we see that the formula underestimates, on average, the point at which this happens and the obtained clustering percentages are conservative.} We show the median of the escape momenta for all halos as the black dash-dotted line, with the 68\% and 95\% containment regions as the light and dark gray shaded area, respectively.

\begin{figure}[ht!]
    \centering
    \includegraphics[width=.8\textwidth]{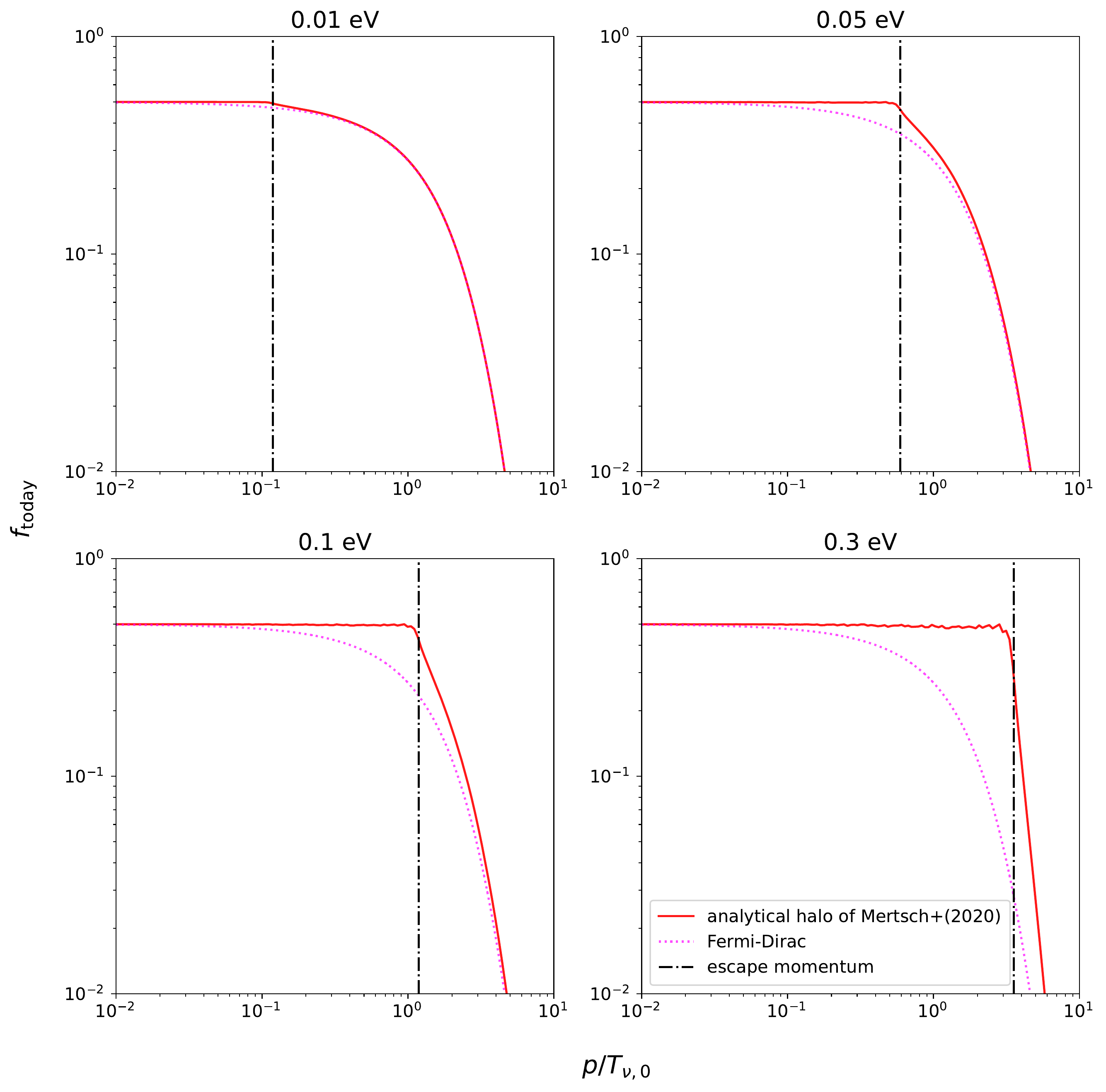}
    \caption{Momentum distributions of today for 4 different neutrino masses using the analytical approach and the same halo parameters as in~\cite{Mertsch:2019qjv}. The (unaltered) Fermi-Dirac distribution is depicted by the dotted magenta line. The black dash-dotted line represents the escape momentum (see text for more details).}
    \label{fig:phase_space_analytical}
\end{figure}

The second effect seems to emerge only with a simulation framework capable of accounting for halos with asymmetric DM distributions. We could not recover this effect with the analytical approach. This can be seen in figure~\ref{fig:phase_space_analytical}, where we show the momentum distribution generated by the analytical approach, using the same halo parameters as in~\cite{Mertsch:2019qjv}. It seems that the DM distributions of the box halos, as well as their more chaotic redshift evolution, cause the less energetic relic neutrinos to diffuse and reduce their likelyhood of reaching Earth. Only above a certain threshold, dependent on the halo characteristics, is the occupation number able to rise and saturate. Bounded from above and below, only relic neutrinos with momenta in the resulting ``window'' are able to significantly cluster and make up the dominant contribution of the local number density. The window gets narrower for higher neutrino masses as more lower momentum states get depleted. This might explain why the blue band in figure~\ref{fig:overdensity_band} starts to slightly dip below the other curves.\footnote{This was also partly shown in~\cite{Mertsch:2019qjv} already, where the inclusion of a Virgo Cluster-like object in their simulations caused the overdensity to slightly decrease for the heavier neutrino masses.} This phenomenon appears to be attributed to the asymmetric nature of the box halos. 
We also approximate the percentage of (fully) clustered neutrinos today,\footnote{A recent study has investigated the survival chance of a potential lepton asymmetry in the CNB~\cite{Ruchayskiy:2022eog}. For this, they computed the bound fraction of relic neutrinos analytically (adjusted with the clustering factors of~\cite{Mertsch:2019qjv}) and found them to be ${\sim}15\%$ for their highest considered neutrino mass of $0.1$ eV.} by integrating the curves in figure~\ref{fig:phase_space_numerical_log} up to the escape momentum thresholds. More specifically, the percentages are obtained with
\begin{equation}
    P(p_\mathrm{esc}) = \frac{\int_{0}^{p_\mathrm{esc}} dp \, p^2 \, f_\mathrm{today}(p)}{\int_{0}^{\infty} dp \, p^2 \, f_\mathrm{today}(p)},
\end{equation}
which yields clustering percentages of $(0.01,0.7,3.9,16.9)\%$ for $(0.01,0.05,0.1,0.3)$ eV neutrinos.

\section{Discussion}
\label{sec:discussion}

Before we conclude, we expand our discussion regarding several points mentioned throughout the previous sections. These include explanations on certain assumptions we made in our simulations, as well as additional interpretations and potential consequences of our findings. 

\paragraph{Cosmological N-body simulations.} The 25 Mpc simulation box employed here did not include neutrinos, which impact the evolution of structure via a suppression of power on scales below the neutrino free-streaming scale of $k_\mathrm{fs} = 0.04 \, h \, \mathrm{Mpc^{-1}} (1+z)^{-1} (\sum m_\nu/ 58 \, \mathrm{meV})$ (see e.g.~\cite{Green:2021xzn}). The suppression can reach ${\sim}20\%$ for $0.3$ eV neutrinos at comoving wavenumbers of ${\sim}1$ $h$/Mpc (corresponding to $\mathcal{O}(10)$ Mpc physical scales today)~\cite{Yoshikawa:2020ehd}. A 25 Mpc simulation box including neutrinos would therefore differ in the structure of its content from a CDM-only box, especially on these scales. However, since we focussed on the impact of the DM distribution (and its evolution) of single MW-type halos on the CNB, we did not extract DM structures from the box beyond these smaller scales, where this suppression of power becomes significant. Our simulation approach is therefore valid on these scales, but one must take caution when conducting simulations with large-scale structure in future continuations of this work.

\paragraph{Relative CNB velocity.} Our simulations were essentially conducted in the CNB rest frame. No relative velocity component was added when initializing the relic neutrinos. Naively, we would expect no significant change to our results for the local overdensities since the loss of flux from one side due to the relative motion should be compensated by the similar gain on the antipodal side. Furthermore, there is an accompanying dipole moment in the CNB anisotropies as calculated in \cite{Lisanti:2014pqa} due to the motion of Earth, but they find this dipole anisotropy to be small, with the percentual difference from the average number density being on the order of ${\sim}0.1\%$. Therefore, the (visible) effect on our anisotropy sky maps should be negligible as well.

\paragraph{Mass \& flavor eigenstates.} We simulated the relic neutrinos as propagating mass eigenstates. Future CNB detectors, like PTOLEMY, will be sensitive to electron flavor neutrinos. The traversed paths of relic neutrinos are on the Mpc scale and the decoherence length of neutrino wavepackets is microscopic (see e.g.~\cite{Akhmedov:2017xxm}). We would therefore expect the CNB to be well mixed and the flavor abundance to be simply given by the measured mixing angles, making our results straightforward to translate to, e.g., $\nu_e$-overdensities. The steps for this are: (i) choose a minimal neutrino mass to establish the three neutrino masses (all must be within our considered neutrino mass range) (ii) take the corresponding number densities ($n_{\nu_i}$) as displayed in figure~\ref{fig:overdensity_band} for these three masses and (iii) convert to local electron neutrino number density via
$n_{\nu_e} = \sum_{i=1,2,3} | U_{ei} |^2 n_{\nu_i}$, where $| U_{ei} |^2$ are the squared mixing (Pontecorvo-Maki–Nakagawa–Sakata or PMNS) matrix elements for the electron flavor (see e.g.~\cite{Esteban:2020cvm}). For the lower limit of number densities possible with our result, i.e. a lighest neutrino mass of $0.01$ eV and normal ordering, we obtain $n_{\nu_e} \approx 56.1 \, \mathrm{cm}^{-3}$. This is similar to the $\Lambda$CDM no-clustering value of ${\sim}56 \, \mathrm{cm}^{-3}$. For a minimal neutrino mass of $0.1$ eV and inverted ordering we obtain $n_{\nu_e} \approx 62.7 \, \mathrm{cm}^{-3}$, which corresponds to a ${\sim}12\%$ increase, and so on.\footnote{In these examples we have divided the value of figure~\ref{fig:overdensity_band} by 2, such that we do not count anti-neutrinos. For further calculations to obtain the capture rate in PTOLEMY, see e.g.~\cite{PTOLEMY:2019hkd,Bauer:2022lri}.}

\paragraph{CNB anisotropies.} Previous works investigating CNB anisotropies~\cite{Michney:2006mk,Hannestad:2009xu,Lisanti:2014pqa,Tully:2021key,Tully:2022erg,Liao:2023zem} have done so, to the best of our knowledge, without the use of full N-body simulations. We showed that the N-1-body methodology (with N-body simulations as external input) is well-suited for computing anisotropy sky maps and should be useful to improve our understanding on the impacts of (large-scale) structure on CNB anisotropies. This could be especially useful for future experiments sensitive to the directions of relic neutrinos. Proposed ideas for this include PTOLEMY with a polarized tritium target or the novel diffraction grating experiment of~\cite{Arvanitaki:2022oby,Arvanitaki:2023fij}. In ref.~\cite{Lisanti:2014pqa}, the authors concluded that even the optimistic case of ${\sim}100$ g tritium in a PTOLEMY-like detector is not enough target mass to observe the (dipole) CNB anisotropies and nonstandard neutrino physics would be required for significant anisotropies. We showed that even with standard physics large anisotropies are possible for neutrino masses of $\gtrsim 0.05$ eV. Since our sky maps predict number density differences up to factors of ${\sim}5$ for different directions, it entails the question if this could improve observability for such future experiments in a meaningful way and would require further investigation.

\paragraph{Phase-Space alterations.} As discussed in section~\ref{sec:results_phase_space}, the depletion of lower momentum states together with the upper bound given by the escape momentum creates a ``window'', where relic neutrinos can be more readily captured. The neutrinos with momenta in this window make up the majority of the local number density. This could have important implications on the detecability of the CNB, especially if the occupation number depletion is significant enough and affects a large range of momenta. With the addition of large-scale structure beyond the MW-like halos we considered in our simulations, this depletion could become more dramatic. More massive DM halos further away and DM filaments in between halos could additively diffuse and trap even more energetic relic neutrinos. How this will ultimately affect the capture rate for PTOLEMY-like experiments and how it competes or is degenerate with other effects deserves in our opinion further investigation.

\paragraph{Helicity composition.} All the gravitational effects described in this work will also alter the helicity composition of the CNB, as they can cause the direction of the neutrino momentum to change. There have been several works investigating the helicity flipping rate due to gravitational potentials and the resulting changes in the helicty composition of the CNB today~\cite{Hernandez-Molinero:2022zoo,Hernandez-Molinero:2023jes,Ruchayskiy:2022eog,Baym:2021ksj}. The amount of fully clustered neutrinos is a proxy for the percentage of the CNB, which has equal amounts of left- and right-helical neutrinos. As we showed in section~\ref{sec:results_phase_space}, this percentage can get as large as ${\sim}17\%$ for the heavier neutrinos. The resulting helicity redistribution due to gravity can affect the capture rate in PTOLEMY-like experiments~\cite{Long:2014zva,Hernandez-Molinero:2022zoo,Hernandez-Molinero:2023jes}, and also erases any (primordial) lepton asymmetries ingrained in the CNB.

\section{Conclusions and Outlook}
\label{sec:conclusions_and_outlook}

In this study, we have demonstrated that the conventional paradigm of gravitational clustering for relic neutrinos is notably influenced when we model local gravitational potentials on the basis of asymmetric DM distributions found in state-of-the-art N-body simulations, contrasting with prior analytical approaches. The key findings of this work can be itemized as follows:

\begin{itemize}
    \item We looked at the predicted local relic neutrino number densities from two different methodologies (section~\ref{sec:methods}), and presented them in figure~\ref{fig:overdensity_band}. We find good agreement between the methods, if the underlying parameters describing the halos are similar.
    \item We find a general correlation between overdensities and virial mass of the halos, while being independent of the concentration and the initial distance of the starting cell in our simulations (figure~\ref{fig:2D_params_box_halos_0.3eV}).
    \item The predicted number densities of the anlytical approach do not vary considerably across the sky, resulting in sub-percent level differences at most (figure~\ref{fig:All_sky_maps_analytical}). The clustering factors are always $> 1$ for all neutrino masses.
    \item When switching to the simulation methodology developed in this work, which accounts for possible DM substructures and inhomogeneities in the underlying DM distribution of the halos, regions of over- \emph{and} underdensity are now possible (figures~\ref{fig:All_sky_maps_benchmark_halo}, \ref{fig:All_sky_maps_halo13} and \ref{fig:All_sky_maps_halo14}). The values for number densities across the sky (across all halos) are in the range of $n_\nu \approx (0.47$--$2.34) \times n_{\nu,0}$ and individual maps can differ up to factors of ${\sim}5$ between their minimum and maximum value. Apart from the benchmark simulation, which displays a clear anti-correlation of number densities with the projected DM content, the correlations for the box halos show no preference either way (figure~\ref{fig:all_power_spectra}), but there do exist halo configurations resulting in correlations and anti-correlations.
    \item The momentum distributions of the relic neutrinos today have been significantly altered under the influence of the evolving gravitational potentials the neutrinos encountered (figure~\ref{fig:phase_space_numerical_log}). Contrary to the predictions from the analytical approach (figure~\ref{fig:phase_space_analytical}), not all momentum states below the gravitational escape threshold are equally likely to reach Earth. The asymmetric DM distributions of the halos in our simulations lead to a diffusion-like effect, resulting in the depopulation of the lower momentum states. Together with the upper limit given by the escape momentum, this creates a momentum window in which relic neutrinos are able to cluster more effectively. Hence, this set of neutrinos makes up the dominant contribution to the local number density. This effect gets only more pronounced with a larger neutrino mass. 
    \item Our simulations predict different percentages of fully clustered relic neutrinos depending on the neutrino mass (see section~\ref{sec:results_phase_space}). However, we used a simple escape momentum formula, which seems to underestimate the point at which the momentum states deplete rapidly for some halos, as visible in figure~\ref{fig:phase_space_numerical_log}. Therefore, the obtained estimates of $(0.01,0.7,3.9,16.9)\%$ for $(0.01,0.05,0.1,0.3)$ eV neutrinos are quite conservative.
\end{itemize}

\noindent A natural extension of this work would entail the use of external N-body simulations that reflect our local environment more realistically, as achieved in \cite{Sawala:2021npe, McAlpine:2022art}. The inclusion of baryonic structures is also a logical step forward to further enhance realism. However, it remains unclear how the addition of more structure, beyond gravitationally bound structures as considered here, would affect the overall local abundance and anisotropies when using our simulation approach. We recall, that simply switching from the methodology in~\cite{Mertsch:2019qjv} to ours resulted in markedly different anisotropy sky maps (figure~\ref{fig:All_sky_maps_analytical} vs. figure~\ref{fig:All_sky_maps_benchmark_halo}), especially creating underdense regions where the DM concentration was higher. Thus, the inclusion of more structure does not neccesarily have to entail an overall increase in the local overdensity, as they could trap or diffuse relic neutrinos as discussed in section~\ref{sec:results_phase_space}.

In summary, it seems that dedicated numerical simulations accounting for non-linear structure formation make it possible to inspect the CNB in a novel way. We presented such a simulation framework that traces the evolution of relic neutrinos to compute local number densities and anisotropies of the CNB, revealing predictions that differ from those obtained with linear theory and analytical methods. To fulfill the ambitious objective of unveiling the early Universe using future CNB observatories, especially those capable of discerning different directions of the incoming neutrinos, it will be crucial to understand the CNB's properties and history through simulations.

\acknowledgments

{\intextsep=0pt
\begin{wrapfigure}{R}{0.125\textwidth}
\includegraphics[width=0.125\textwidth]{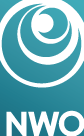}
\end{wrapfigure}
We thank Phillip Mertsch for kindly answering questions about his work and Chris Tully for fruitful discussions. The simulations for this work were performed on the Snellius Computing Clusters at SURFsara. This publication is part of the project ``One second after the Big Bang'' NWA.1292.19.231 which is financed by the Dutch Research Council (NWO). CC acknowledges the support of the Dutch Research Council (NWO Veni 192.020). SA was partly supported by MEXT KAKENHI Grant Numbers, JP20H05850 and JP20H05861.\par}

\appendix

\section{Engineered NFW Halo}
\label{app:NFW_halo}

Here we describe the construction of the NFW-based halo used for our benchmark simulations (see section~\ref{sec:results}). The halo is made up of individual DM particles, which have the same mass as the particles in the 25 Mpc cosmological simulation box we employed. First we sample radii from a NFW distribution~\cite{Navarro:1995iw}
\begin{equation}
\rho(r) =  \frac{\rho_0}{\frac{r}{R_s} \left( \frac{r}{R_s}+1 \right)^2} \Theta(r - R_{200}),
\label{eq:NFW_profile}
\end{equation}
which we truncate at the virial radius $R_{200}$ with the Heaviside function $\Theta$, such that we can obtain the finite virial mass via
\begin{equation}
M_{200} = M(r = R_{200}) = 4 \pi \rho_0 R_s^3 f(r)
\label{eq:virial_mass}
\end{equation}
where the helper function $f(r)$ is
\begin{equation}
f(r) = \ln(1+(r/R_s)) - \frac{r/R_s}{1+(r/R_s)}.
\label{eq:helper_function}
\end{equation}
\begin{figure}[t!]
    \centering
    \includegraphics[width=.6\textwidth]{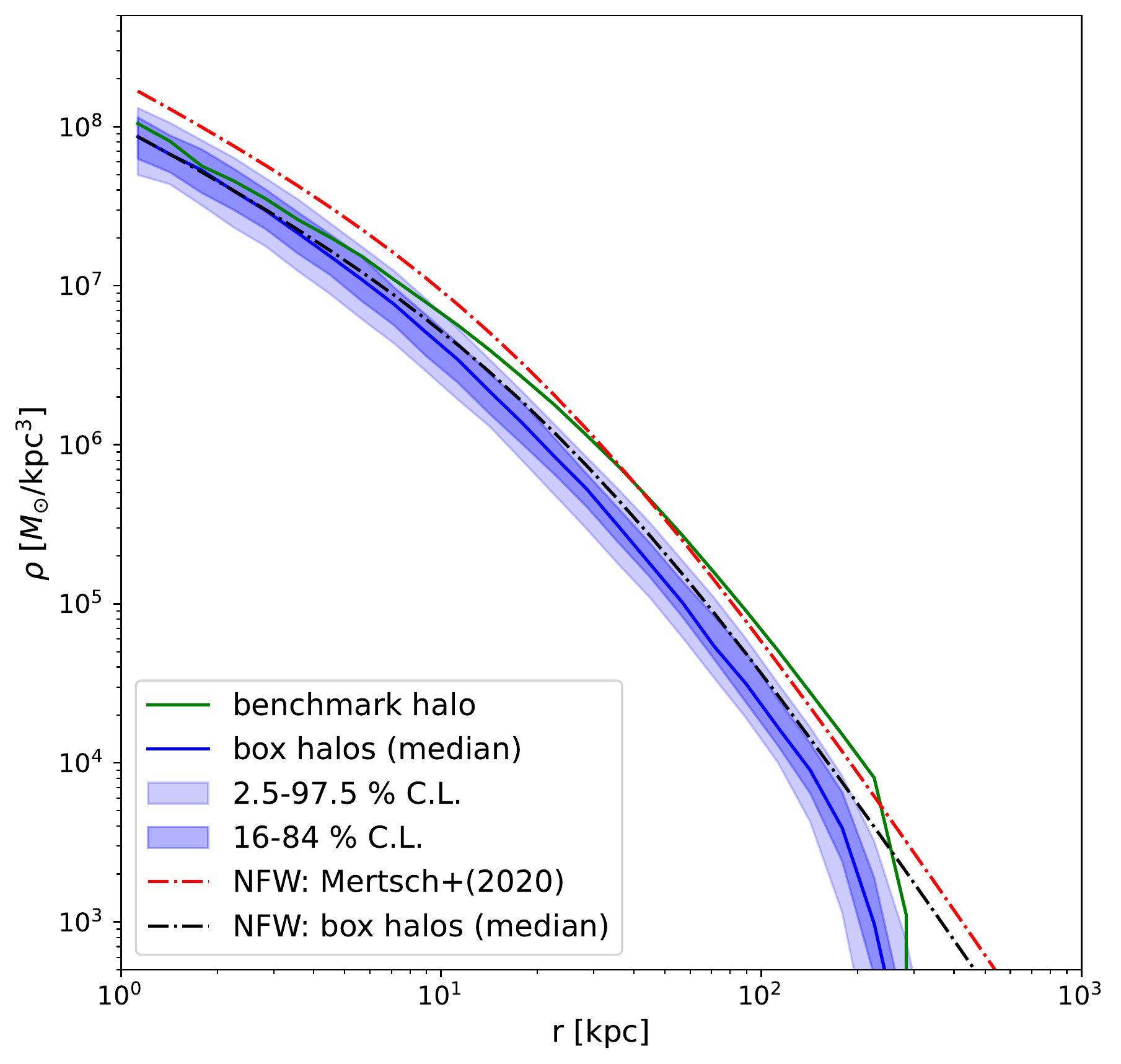}
    \caption{Radial density profile for our halo objects. The blue curve represents the median densities of our box halo sample, with the $68\%$ and $95\%$ coverage band as the dark and light blue shaded region, respectively. The green curve represents the benchmark halo. For comparison and validation purposes, the dash-dotted red and black curves show the NFW profiles using the NFW parameters $(\rho_0, R_s)$ from~\cite{Mertsch:2019qjv} and from of our box halo sample (medians), respectively.}
    \label{fig:density_profiles}
\end{figure}

The function $y = M(r \leq R_{200})/M_{200}$, i.e. $y = f(r)/f(R_{200})$, will then give us a value between $0$ and $1$ for a radius between $0$ and $R_{200}$. We can now uniformly sample values between $0$ and $1$ and get a distribution of radii using the inverse of this function.\footnote{We used the python library \texttt{pynverse} to numerically find and evaluate the inverse function.} However, we need to evolve the scale radius of the halo with the redshift. For this we used the \texttt{COMMAH} library \cite{Correa:2014xma,Correa:2015dva,Correa:2015kia} in python. For a given virial mass at redshift $0$ it can return the virial mass, $M_{200}(z)$, and concentration, $c_{200}(z)$, for an array of redshifts. We then obtain the virial radius with
\begin{equation}
R_{200}(z) = \left( \frac{M_{200}(z)}{200 \rho_{crit}(z) \frac{4}{3} \pi} \right)^{\frac{1}{3}},
\end{equation}
where $\rho_{crit}(z)$ is the critical density of the Universe at redshift $z$. The scale radius is $R_s = R_{200}/c_{200}$. We can now sample a number of DM particles, i.e. a number of radii, equal to $\lfloor M_{200}(z)/m_{DM} \rfloor$. We then isotropically distribute these radii to cartesian coordinates with
\begin{equation}
\begin{aligned}
    x &= r \cos(\phi)\sqrt{1 - \cos^2(\theta)}\\
    y &= r \sin(\phi)\sqrt{1 - \cos^2(\theta)}\\
    z &= r \cos(\theta),
\end{aligned}
\end{equation}
where the azimuthal angles are uniformly sampled between $0$ and $2\pi$ and the cosines of the polar angles are uniformly sampled between $-1$ and $1$. We show the density profiles of all objects considered in our simulations in figure~\ref{fig:density_profiles}.





\providecommand{\href}[2]{#2}\begingroup\raggedright\endgroup
\end{document}